\documentclass[aps,12pt,tightenlines,amsmath,amssymb]{revtex4}
\usepackage{graphics}
\usepackage{dcolumn}
\usepackage{bm}
\usepackage{epsfig}

\usepackage{color}

\newcommand{\labeld}[1]{ }

\font\capseight=cmcsc8

\newcommand{\be}{\begin{equation}}\newcommand{\ee}{\end{equation}}
\def\Eqref#1{Eq.~(\ref{#1})}
\def\matlab{{\capseight MATLAB}\textsuperscript{\tiny\textregistered}}
\def\Rorig{R_{\mathrm{orig}}}

\def\noteconditions{The only conditions on $R$ are that it be stochastic and irreducible. It need not satisfy detailed balance. In practice, for Fig.\ \ref{f.01} the matrices were created in the following way. Each element was independently chosen from a uniform distribution on $[0,1]$\@. The diagonal was set to zero. Then each column was divided by the column sum. I then checked if the matrix was irreducible. If not, another matrix was created in the same way, until I obtained an irreducible matrix. (In practice this was generally successful on the first try.)}

\def\noteparameters{The parameters used were $R=0.25$, $Q=5$, $\ell=1$, $a\approx1.0356$, and $b\approx0.0230$\@. The arena for the walk is the square of side $2L$ and is discretized as 25 points in each direction. Below we introduce an additional contribution to the potential, for which the parameter values are $T=4$ and $\alpha=0.15$\@.}

\def\noteear{An example of ``playing it by ear'' could occur if one had an a priori idea of what the transition probabilities ought to be, as in the TN model.}

\begin{document}
\title{Transition matrix from a random walk}

\author{Lawrence S. Schulman}
\affiliation{Physics Department, Clarkson University, Potsdam, New York 13699-5820, USA}
\affiliation{Max Planck Institute for the Physics of Complex Systems, N\"othnitzer Str.\ 38, D-01187 Dresden, Germany}
\email{schulman<at>clarkson.edu}

\date{\today}
\begin{abstract} Given a random walk a method is presented to produce a matrix of transition probabilities that is consistent with that random walk. The method is a kind of reverse application of the usual ergodicity and is tested by using a transition matrix to produce a path and then using that path to create the estimate. The two matrices and their predictions are then compared. A variety of situations test the method, random matrices, metastable configurations (for which ergodicity often does not apply) and explicit violation of detailed balance.
\end{abstract}

\maketitle

\section{Introduction\label{s.intro}\labeld{s.intro}}

It can happen that one knows a specific path of a stochastic process but not the transition probabilities that generated it. Or perhaps one can, in principle, know each transition probability but the state space is so huge that one does not have practical control over the underlying transition probabilities. Suppose that in the face of this difficulty one would nevertheless want properties of the matrix (of transition probabilities) that generated the particular path. In this note we present a method for estimating this matrix from a particular path. What we will present is an estimate: you don't know what you don't know. This means that the path itself does not have information about regions of the space that are not entered (as in metastability). What we will obtain is properties of the actual path; if there is another possible path, perhaps in another region of the state space, this method will not reveal it.

What good is such partial information? I will give examples where recovery of an underlying transition matrix sheds light on the path and the space within which it travels. Our examples will occasionally use the \textit{observable representation} which, because it appears later in this article, is briefly described in Appendix \ref{s.observable}, and concerning which more information is available in \cite{note:visualizing}. Other parts involve the Tangled Nature model and foraging, which are only motivational and for which we only provide citations (\cite{jensen} and \cite{daluz}, respectively). As explained, the \textit{observable representation} provides (among other things) an embedding of the (possibly) abstract coordinate space in a low dimensional Euclidean space---or, more precisely, it embeds that portion sampled by the path. Moreover, this embedding reflects proximity features (in a dynamic sense) of the transition matrix. The examples arose in dealing with both the Tangled Nature model and foraging; in both cases the information provided by the present method allowed an embedding to be constructed and a led to a better understanding of the structure of the underlying spaces. Thus when discussing foraging, the location of the animal's food supplies could be located~\cite{mwthesis}.

In this note I present a way to produce the transition probabilities from the path, even if the state space is in principle infinite. In this method, I focus on those sites actually visited by the random walk, thereby imposing an effective cutoff.

The method used here is a kind of ergodicity. Usually one is interested in temporal averages along some path, but instead---appealing to ergodicity---one actually calculates phase space (or Hilbert space) averages. We are going in the opposite direction. We have the time dependence and wish to deduce from it the ``spatial'' behavior. In this case the space is the set of sites visited on the random walk and ``behavior'' amounts to knowing the transition probability among the sites. It is amusing that the method can be applied to metastable walks, where ergodicity is patently absent, with the extraction of useful results for the restricted walk.

\section{The method \label{s.method}\labeld{s.method}}

Suppose we are given a specific random walk, $x(t)$, $t=1,\dots,T$\@. For our purposes the nature of the space in which $x$ takes its values is irrelevant, and we replace it by integers. Thus $x(1)$ is taken to be 1. If $x(2)$ is different it is called 2; otherwise it is also 1. The total state space \textit{of interest in the present construction} is the set of unique values taken by $\{x(t)\,|\,t=1,\dots,T\}$\@. Bear in mind that the actual state space may in principle be infinite (as in the TN model), but this is not a problem for us.

Let the number of unique sites be called $N$, so that the set of transition probabilities is an $N\!\times\!N$ matrix. Let (our forthcoming estimate of) that matrix be called $R(a,b)=\Pr(a\leftarrow b)$, where $a$ and $b$ can be considered to be integers between 1 and $N$\@. We define an intermediate quantity $S(a,b)$ to be the number of times (out of all $T-1$ time steps) that $b$ goes to $a$\@.

This is not yet what is needed---even aside from correcting the normalization (column sums being 1). The reason is that this process may leave the resulting transition matrix reducible. Thus the initial state, $x(1)$, may be an unlikely one, and the system never returns to it. For this reason we impose the artificial condition that $x(T)$ is followed by $x(1)$, so that $S(x(1),x(T))$ is at least 1\@. If $T$ is large this step should not be overly harmful, but there is no guarantee.

Once $S$ is obtained, it is divided by its column sums, producing the desired transition matrix. In \matlab\ notation this is {\tt R=S*diag(1./sum(S))}. The resultant $R$ is a stochastic irreducible matrix.

It may be worth mentioning that this is \textit{not} an exercise in the central limit theorem (although error estimates may use this). On the contrary: I appeal to no theorem. Perhaps there are theorems waiting to be proved, but our present goal is to provide a practical method and to test its applicability in various circumstances. To put it simply: you have information (the path); what other information can you draw from it?

\section{Numerical checks\label{s.numerical}\labeld{s.numerical}}

\noindent\textsf{\small Notation: Throughout the following, $\Rorig$ is the ``original'' transition matrix generating the path. $R$ is the estimate derived from the path, as described above. It is assumed that $\Rorig$ is irreducible.}

\medskip

One way to check this technique is to generate a random walk using a particular matrix of transition probabilities and to use that same random walk to produce a matrix of transitions according to the method described above. The original and the derived matrix are then compared.

\begin{figure}[h]
\includegraphics[height=.3\textheight]{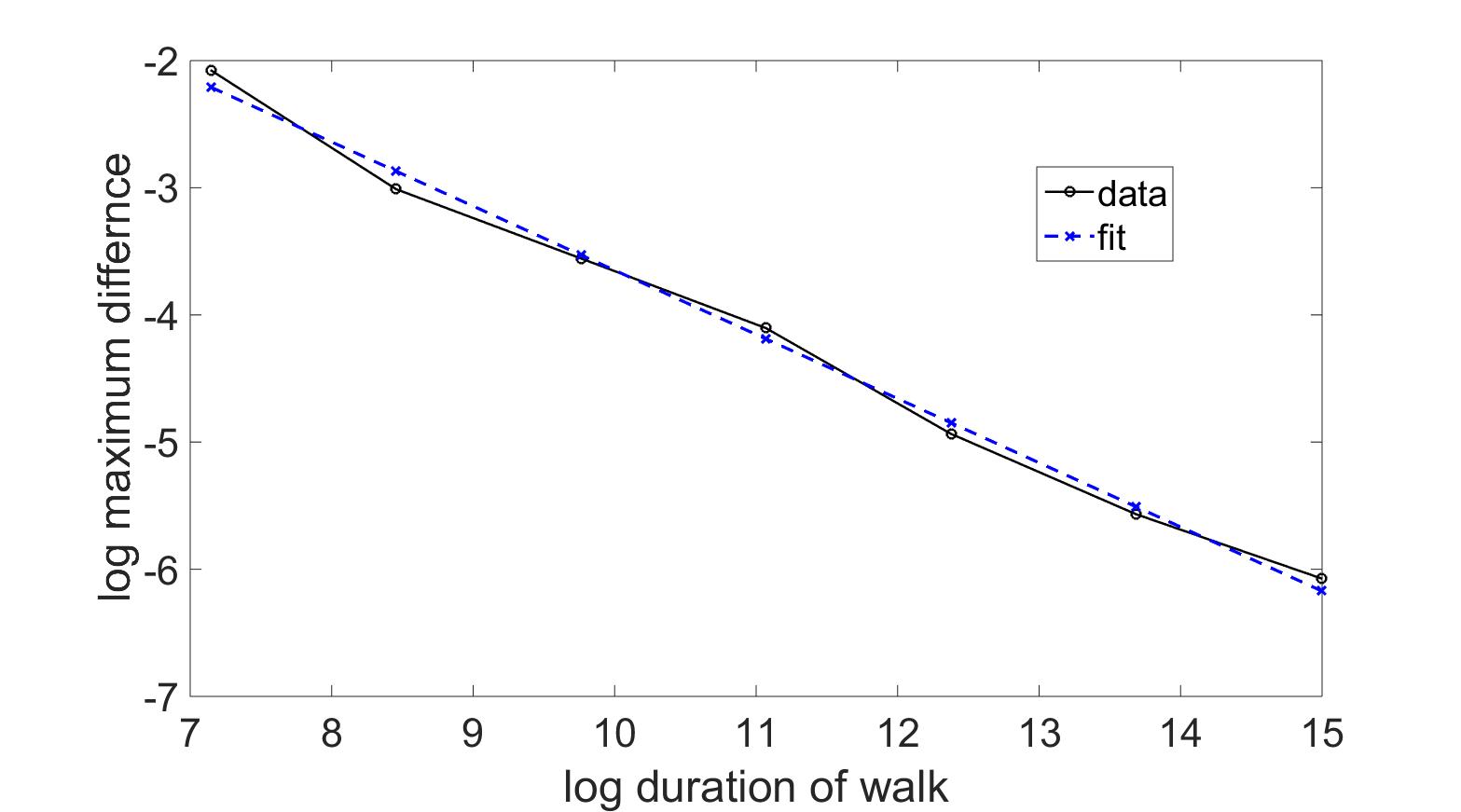}
\caption{Log-log plot of the maximum difference of the original transition matrix and the deduced matrix as a function of path length of the random walk. The dashed line is the fit, described in the text.\label{f.01}\labeld{f.01}}
\end{figure}

We illustrate our results with 25$\times$25 matrices. We began with a matrix $\Rorig$ which generates the random walk, according to its probabilities (see \footnote{\noteconditions} for details of the matrix construction), to obtain some $x(t)$, $t=1,\dots,T$\@.  Then using $x(t)$ and following the method of Sec.\ \ref{s.method} we produced the estimated matrix, $R$\@. Our measure of their difference from one another is the maximum of absolute values of their matrix elements, thus $D(T)\equiv \max|R-\Rorig|$ (where this is a maximum over rows \textit{and} columns (unlike \matlab's convention). In Fig.\ \ref{f.01} we show a plot of $\log D(T)$ vs.\ $\log T$\@. From several such plots we find that $D(T)\sim \mathrm{const}/T^{1/2}$\@. By checking other matrix sizes we have found that the appearance of the square root is not sensitive to~$N$\@. This is reasonable, since the fluctuations are essentially $1/\sqrt{\mathrm{smallest} \, S(a,b)}$ and $S(a,b)$ is in turn proportional to $T$\@.

\begin{figure}[h]
\includegraphics[height=.3\textheight]{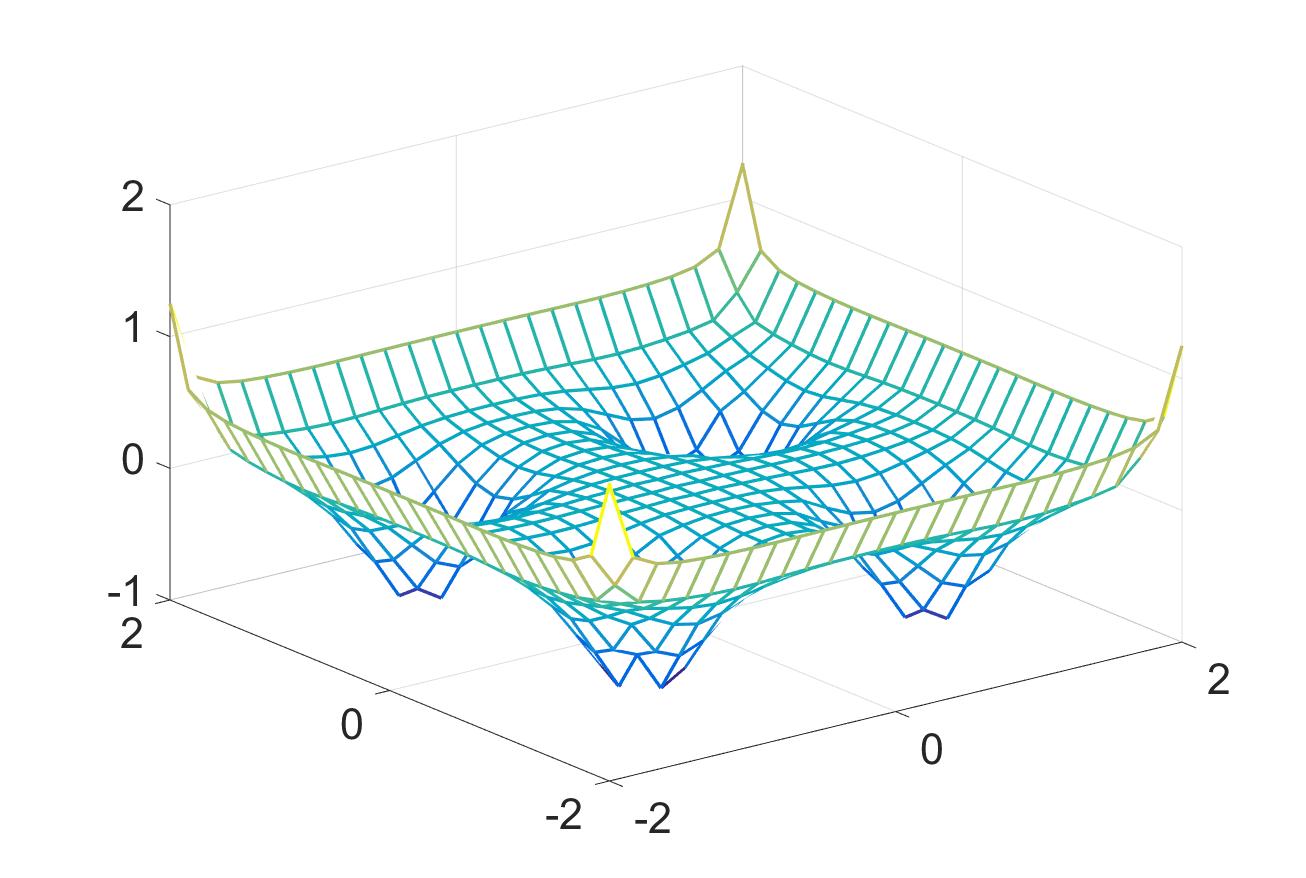}
\caption{Potential for a random walk. There will be a tendency (governed by the inverse temperature $\beta$) for the random walk to spend the most time in the minima. For infinite time or small $\beta$ there will be four preferred states and one could consider the system to be one having four phases with a first order phase transition in going from one to the other. (See App.\ \ref{s.observable} for the concept of first order phase transition in this context.) \label{f.02}\labeld{f.02}}
\end{figure}

It is also of interest to test this method when the random walk has further features, for example, the presence of metastable states. We will also study the currents (shortly to be defined precisely). Consider a random walk in the potential illustrated in Fig.~\ref{f.02}, which has the basic form
\be
V_0=-a\sum \exp\left(-Q(x\mp\ell)^2 - Q(y\mp\ell)^2 \right) +b\left(\exp(Rx^4)+\exp(Ry^4)      \right)
\,,
\label{e.010}
\ee
\labeld{e.010}
for positive constants $a$, $b$, $Q$ and $R$ \footnote{\noteparameters}, and the ``$\,\mp\,$'' sum is over all 4 possibilities. $u\equiv(x,y)$ is defined with each variable on an interval $[-L,L]$ and discretized. The space $X$ is thus the set of pairs $\{u\}=\{(x,y)\}=\{(n_x,n_y)\}/6$, with $-12\le n_x,n_y \le 12$ (with the $n$'s integers). In the picture we have added a small potential, which allows enhanced tunneling between certain pairs of the minima, but not the other pairs. The potential is $W(x,y)=-\alpha\left(\exp(-T(x-\ell)^2)+\exp(-T(x+\ell)^2)\right)$\@. Define $V\equiv V_0+W$\@.

We assume that the walker moves only to nearest neighbor sites. The transition probabilities for this process are trivially known and we again refer to the original matrix as $\Rorig$\@. A walk in this potential satisfies detailed balance and at inverse temperature $\beta$ has the stationary state
\be
p_0(u)=\frac1Z \exp\left(-\beta V(u)\right)\;\quad\hbox{with}\quad Z=\sum_u \exp(-\beta V(u))
\,.
\label{e.020}
\ee
\labeld{e.020}
What is particularly interesting in this case is that for large $\beta$ or for short walks the path can be stuck in a metastable state. The resulting transition matrix will no longer be the same as the entire matrix. Does it have any value, nevertheless?

\begin{figure}[h]
\includegraphics[height=.3\textheight]{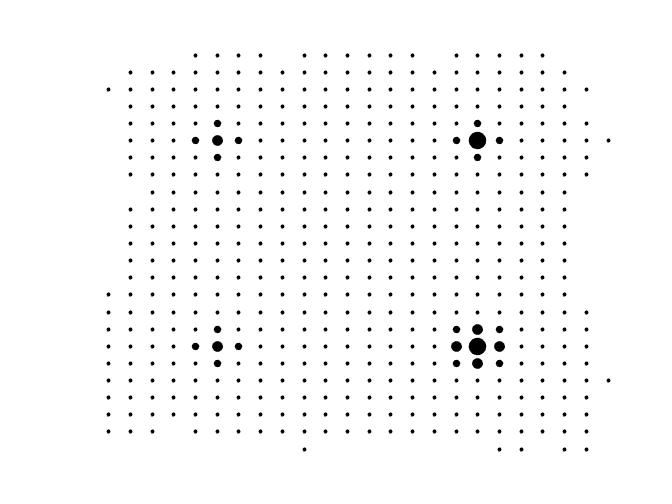}
\caption{Sites visited by the random walk. Each spot is a site in the $x$-$y$ plane of \Eqref{e.010} or Fig.\ \ref{f.02}. The spot size is a monotonic function of the number of visits to that site. Sites never visited do not appear. The initial site for the walk was on the lower right. \label{f.03}\labeld{f.03}}
\end{figure}

\begin{figure}[h]
\includegraphics[height=.3\textheight]{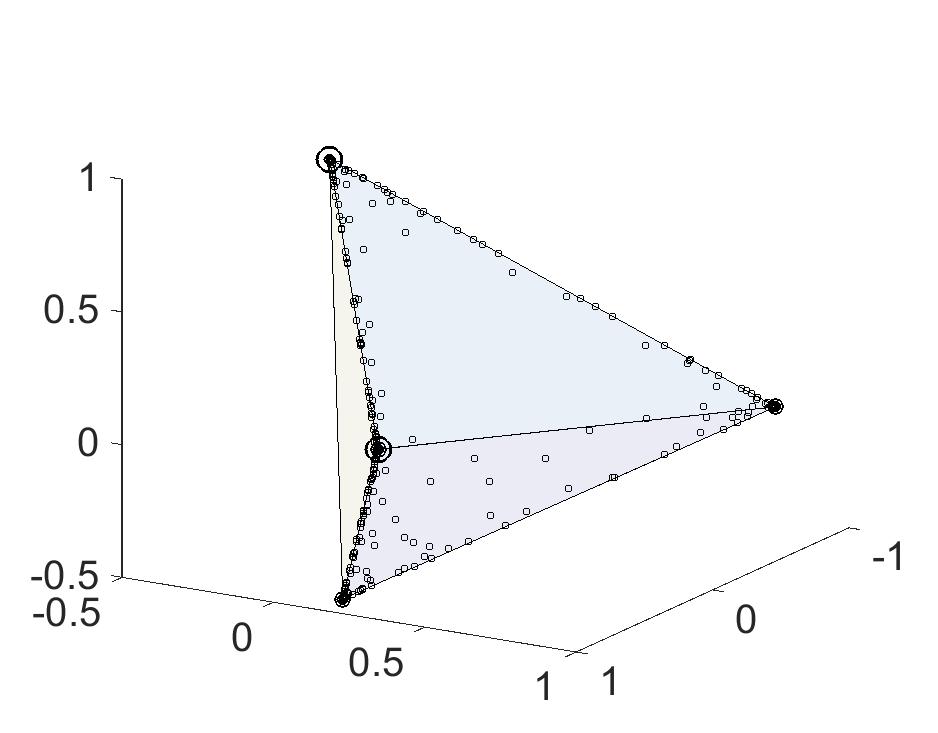}
\caption{Observable representation (OR) for $R$, a clean tetrahedron (cf.\ App.\ \ref{s.observable}), essentially identical to what results from an analysis of $\Rorig$\@. The estimator matrix ($R$) in this case is based on the long path, comprising $10^7$ steps. Symbol size is a monotonically increasing function of probability. It is clear that the bulk of the probability is located at the vertices---as it should be for first order transitions with 4 stable states. \label{f.04}\labeld{f.04}}
\end{figure}

To gain perspective on this issue we first examine the path when the number of steps (of the path, its duration) is rather long. Let the path have $10^7$ steps on a $25\times25$ mesh (so $\Rorig$ is $625\times625$). For $\beta=6$ (and $\min V=-1.27$) the path enters 505 of the 625 possible sites (in a run that uses a particular sequence of ``random'' numbers as selected by the software). One can look at the ground state probability derived from $R$, illustrated in Fig.~\ref{f.03}. In that figure the size of the spot is monotonically related to the stationary distribution of $R$ (which is almost proportional to the number of times a given site is visited). Equally significant is the observable representation image (Fig.\ \ref{f.04}) obtained from $R$, which is a clean tetrahedron (which it should be---cf.\ App.\ \ref{s.observable}). The 4 vertices represent the 4 phases implicit in the potential with the bulk of the probability of the stationary distribution located on the vertices. (For more on the observable representation see App.\ \ref{s.observable} and \cite{note:visualizing, imaging, multiplephases, meanfieldobsrep}.) Now the matrix $R$ can in principle violate detailed balance (and generically does) even though $\Rorig$ does not. However, in this case the violation is insignificant: the largest current matrix element is about $6\times10^{-5}$\@. (The current matrix is defined as $J_{ab}\equiv R_{ab}p_0(b)-R_{ba}p_0(a)$ with $p_0$ the stationary distribution of $R$, $a,b\in X$ and $X$ the system space. Having the current be entirely zero is the same as detailed balance.)

\begin{figure}
\includegraphics[height=.15\textheight]{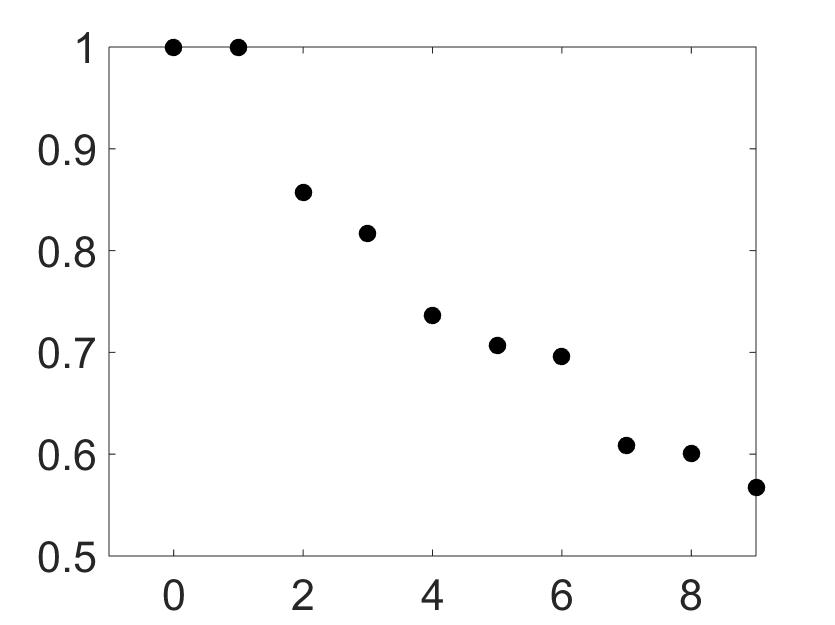}
\caption{Spectrum of $R$ for a shorter path. The ordinate is the value of the eigenvalue, the abscissa the eigenvalue label. A corresponding plot for the the estimator matrix based on the long path would show 4 eigenvalues close to 1. \label{f.05}\labeld{f.05}}
\end{figure}

\begin{figure}
\includegraphics[height=.2\textheight]{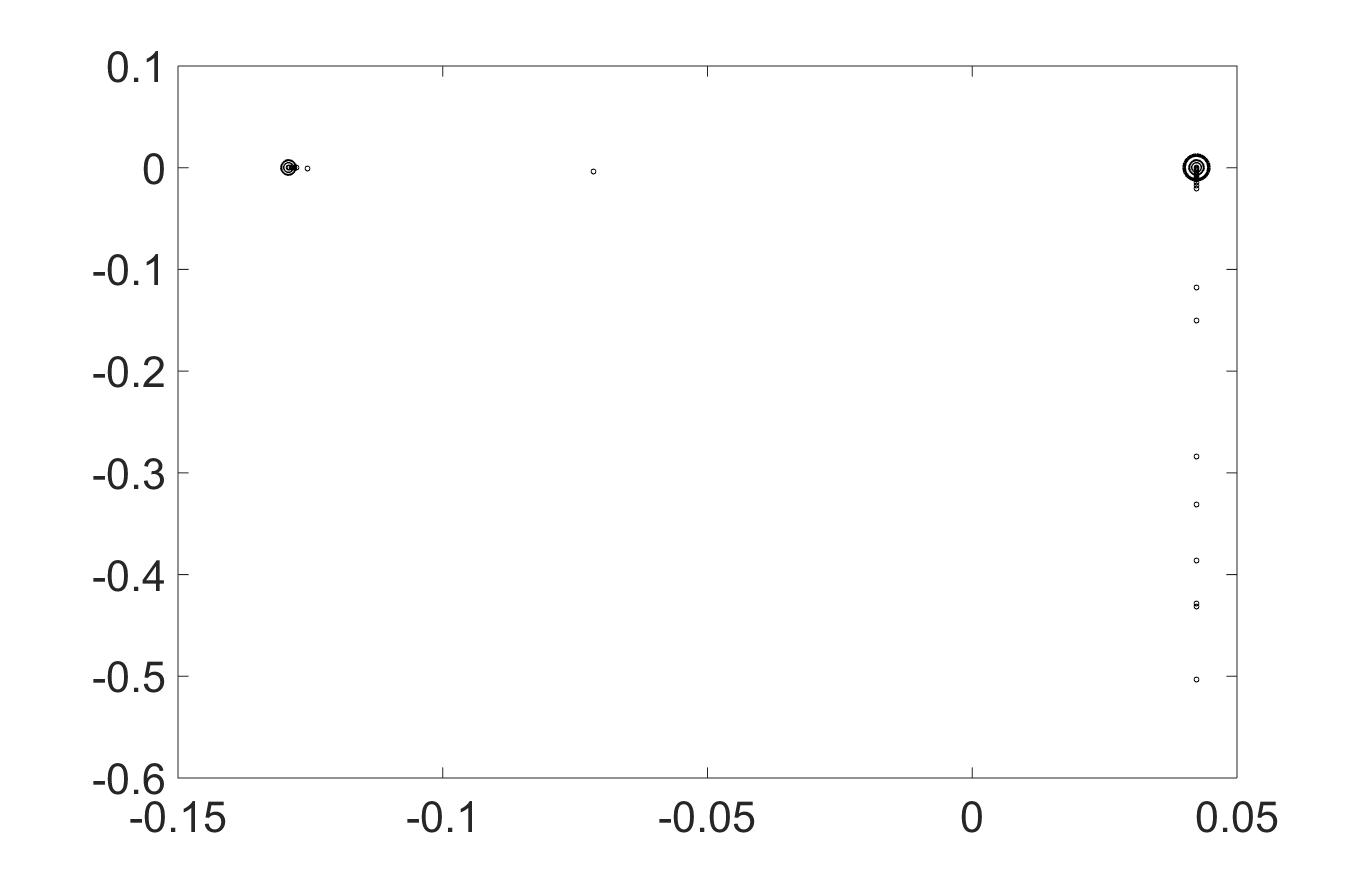}
\caption{Two dimensional observable representation (OR) for this ``phase transition.'' Since the shorter path only samples part of the space the reconstruction of $R$ does not know of other regions of $X$\@. As far as the reconstructed $R$ for this path is concerned there are only two phases. The OR for a two phase system is a one-dimensional line with points having the most weight (highest values of $p_0$) at either end of the line. Since we have plotted a \textit{two}-dimensional figure there is spread along the $A_2$ axis, spread that is consistent with the presence of just two phases. (See \cite{note:visualizing} for further explanaton.) \label{f.06}\labeld{f.06}}
\end{figure}

\begin{figure}
\includegraphics[height=.2\textheight]{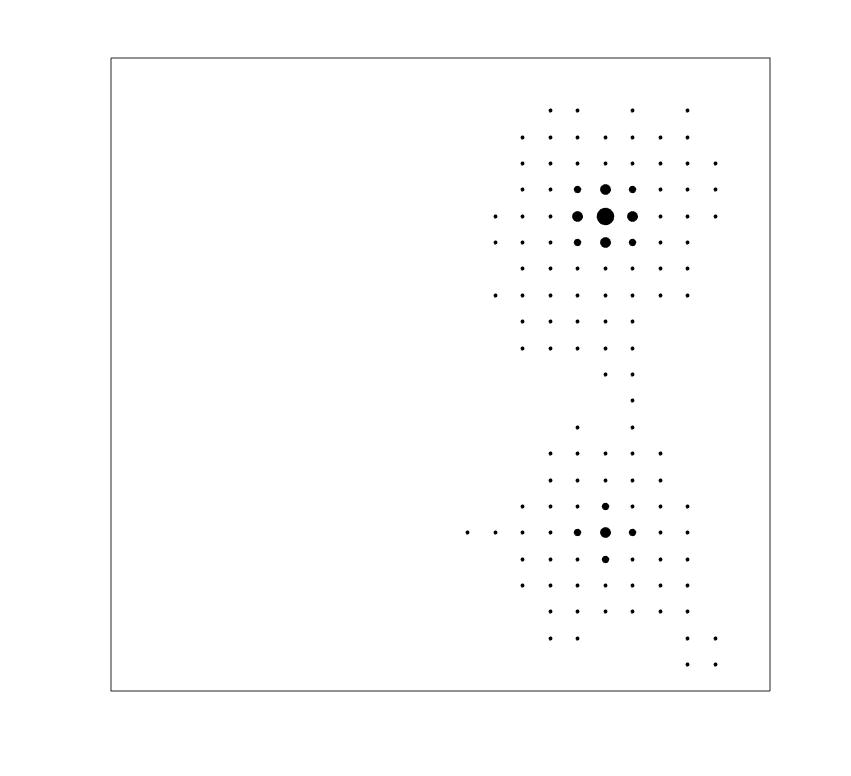}
\caption{Same diagram as Fig.\ \ref{f.03}, but for a different path. Shown are the points visited, with weights, for a shorter path. The boundary of the space is drawn (which is not the case in Fig.\ \ref{f.03}) to emphasize that many points were never visited. \label{f.07}\labeld{f.07}}
\end{figure}

Next we consider paths considerably shorter: $2\times10^5$ steps, also at $\beta=6$\@. Only 126 of the 625 points are visited. Seeing in what way they capture the essence of $R$ requires some familiarity with the OR\@. When the OR gives a clean tetrahedron (i.e., for a longer path or for $\Rorig$), besides the eigenvalue 1 of the transition matrix that always exists for a stochastic matrix, there were three others, all very close to one. In this case $R$ has only one other, as can be seen in an image of the eigenvalues closest to 1, in Fig.~\ref{f.05}. Under these circumstances there is a first order transition between only two states. In principle a one-dimensional figure is all that is needed for the OR, namely the values of $A_1$, the first non-trivial left eigenvector of $R$\@. In Fig.\ \ref{f.06} we plot $A_2$ as well, but the fact that contributions line up along the $A_1$ axis already shows that this single dimension is capturing the essential properties of $R$\@. As compelling final view we show in Fig.\ \ref{f.07} the points visited on this shorter path, together with probability weights. (The term $W$ in the potential has helped fix the asymmetry.)

\begin{figure}
\includegraphics[height=.2\textheight]{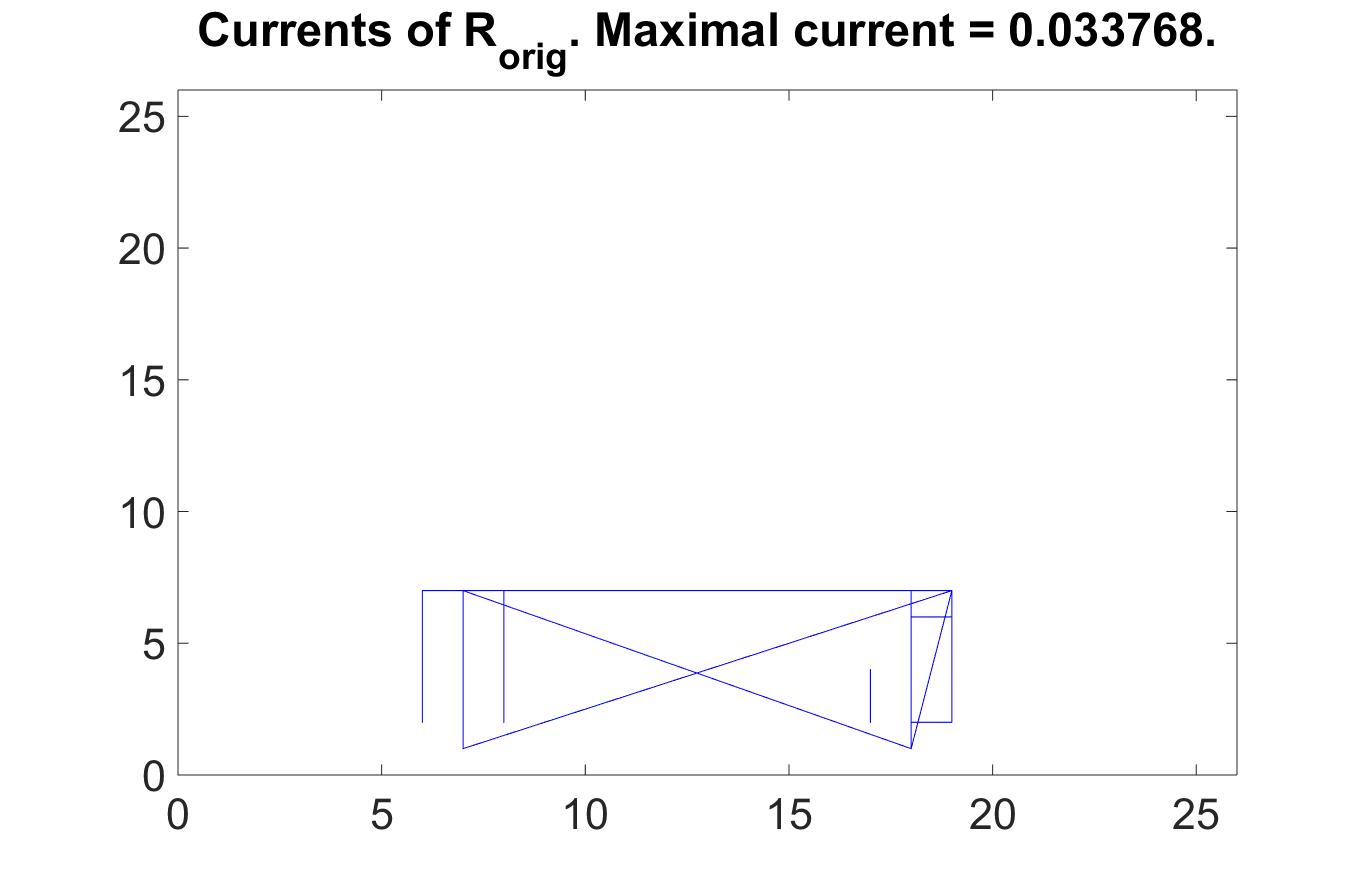}
\includegraphics[height=.2\textheight]{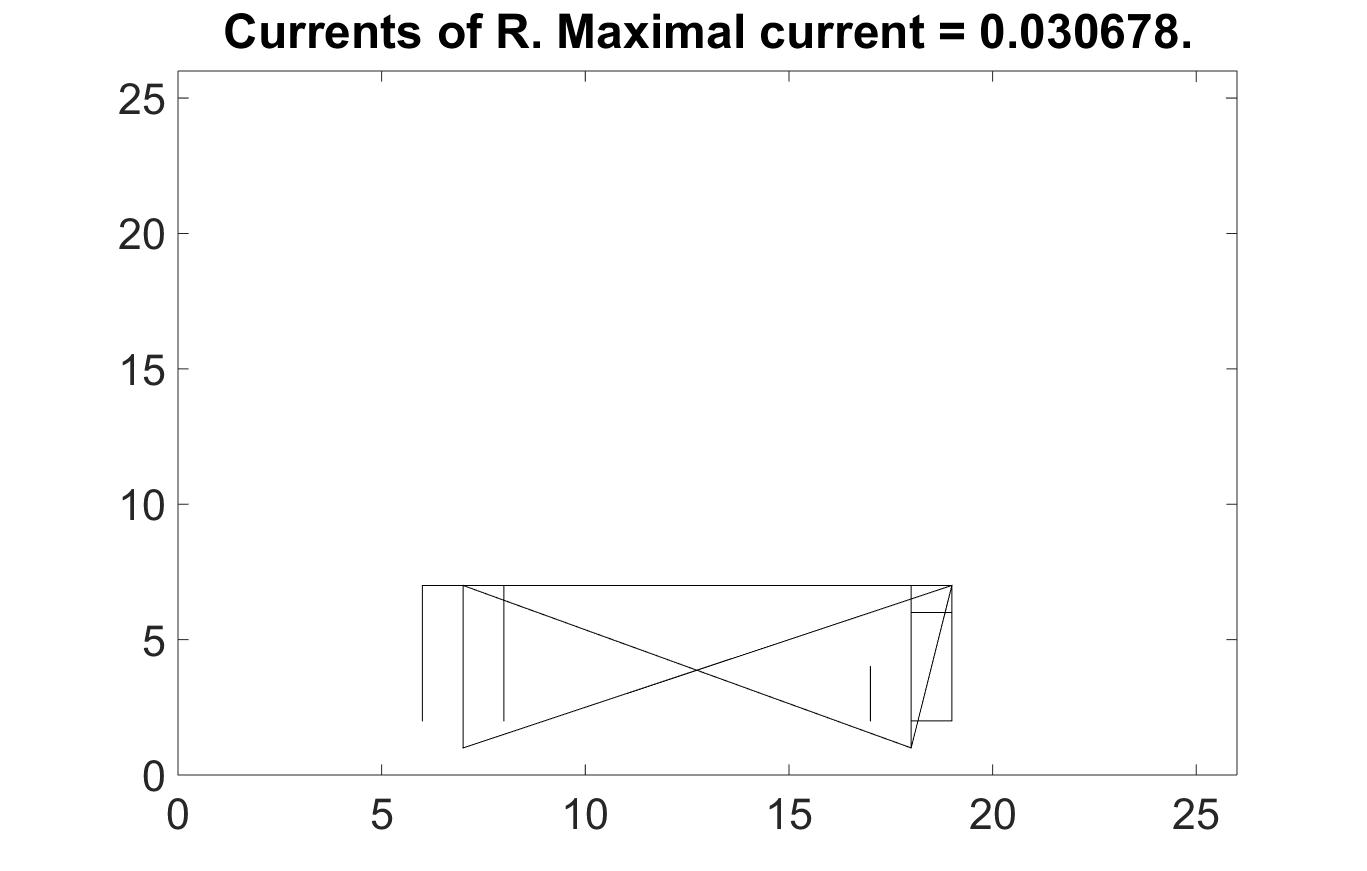}
\caption{Currents for $\Rorig$ and for $R$, the latter involving a long path. A line from a point (say) $(2,3)$ to (say) $(6,8)$ indicates that the matrix element $j_{(2,3)(6,8)}$ is above 5\% of the maximum $j_{ab}$ value. The direction of the current is not indicated.  \label{f.08}\labeld{f.08} What is illustrated is not the signed current, $J_{ab}$, but $j_{ab}\equiv(J_{ab}+|J_{ba}|)/2$\@. Shown is the 25$\times$25 array corresponding to $X$\@. (As indicated earlier, the values in $X$ are pairs from $-2$ to $+2$, spaced by 1/6, $25^2$ points in all; Since the actual $x$ and $y$ values are of no interest, we have simply labeled $X$ as a 25-by-25 array.) }
\end{figure}

\begin{figure}
\includegraphics[height=.2\textheight]{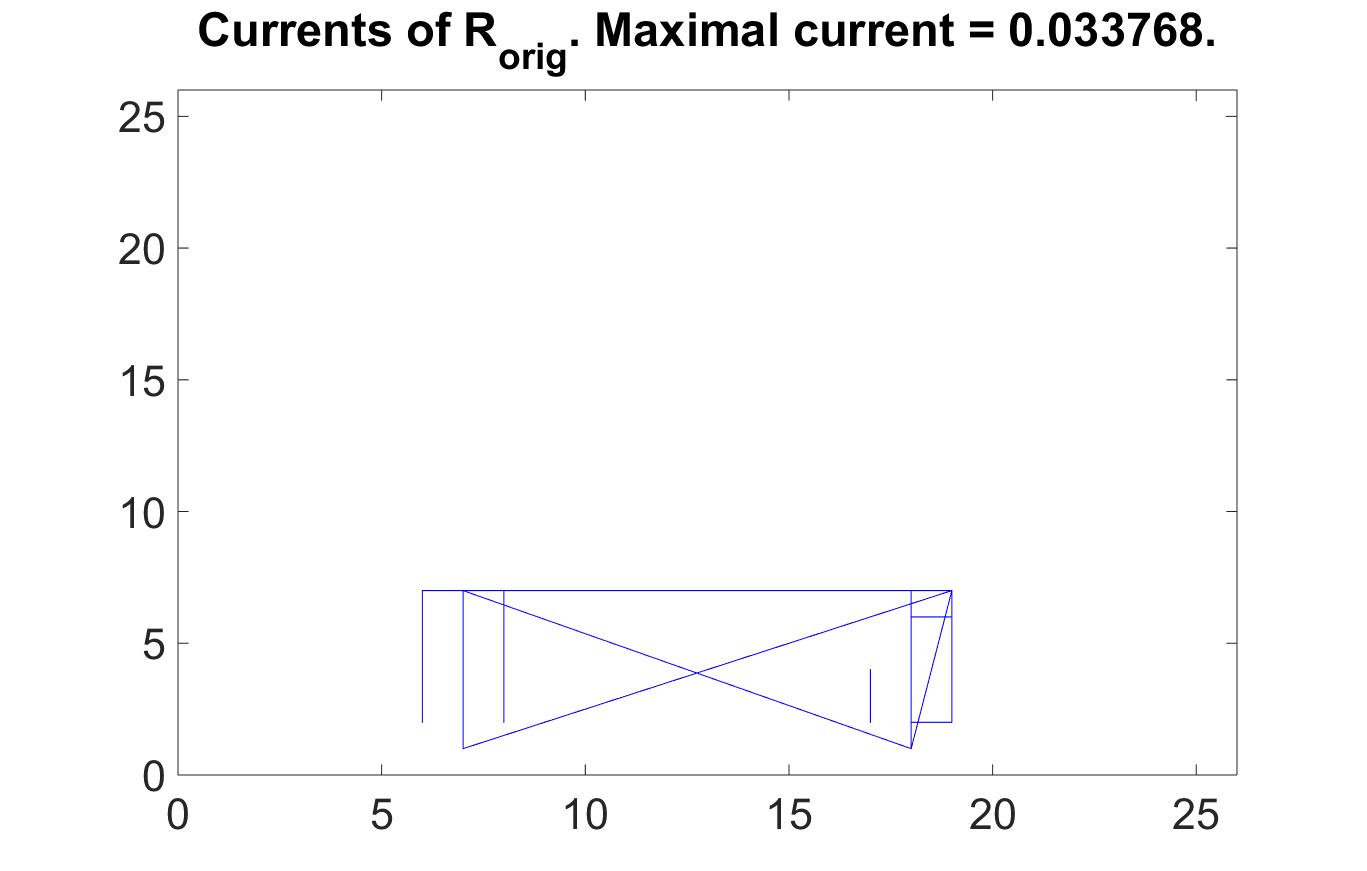}
\includegraphics[height=.2\textheight]{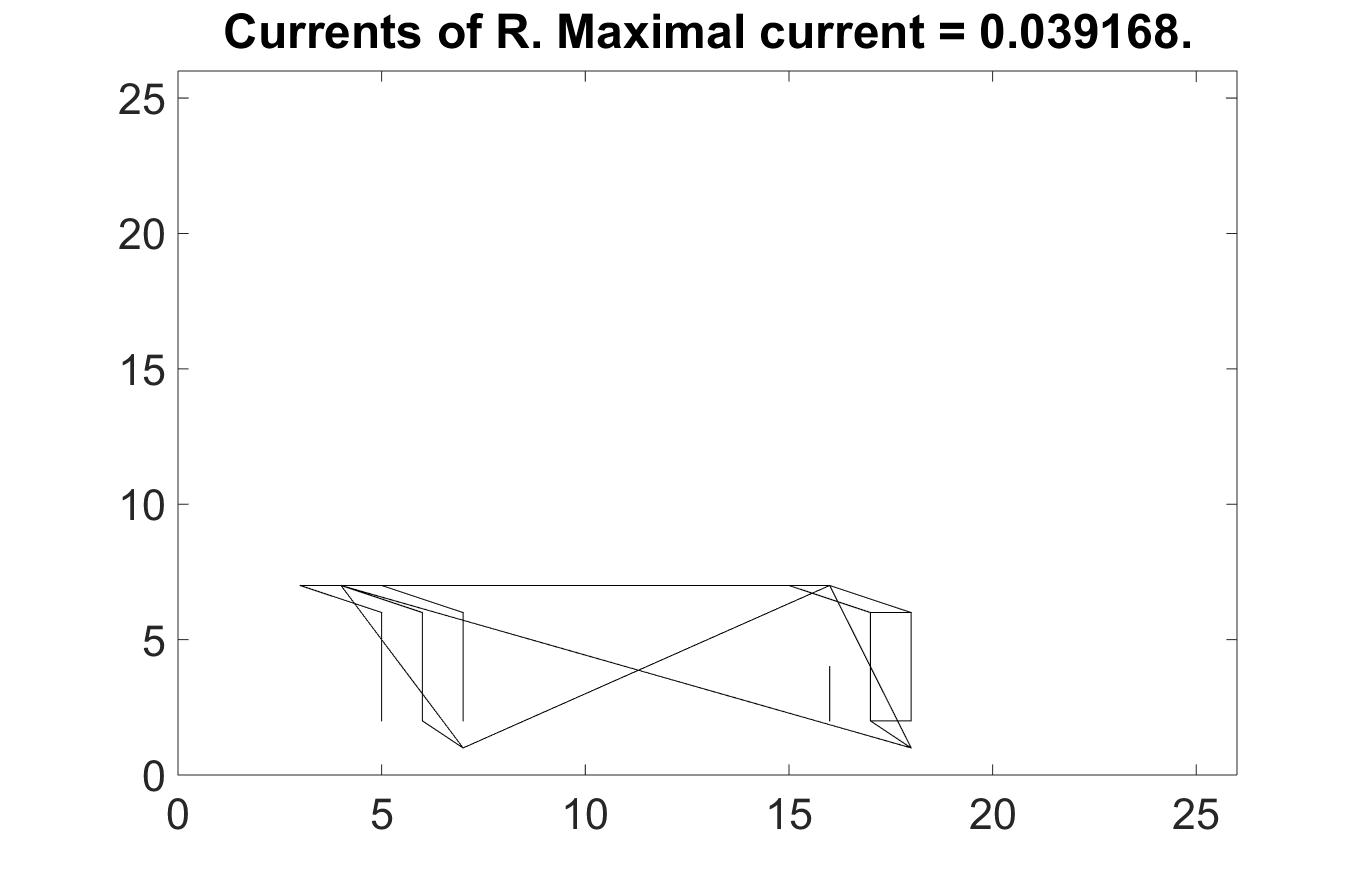}
\caption{Currents when $R$ is based on a shorter path. Again only the positive component of the current is shown. \label{f.09}\labeld{f.09}}
\end{figure}

\begin{figure}
\includegraphics[height=.2\textheight]{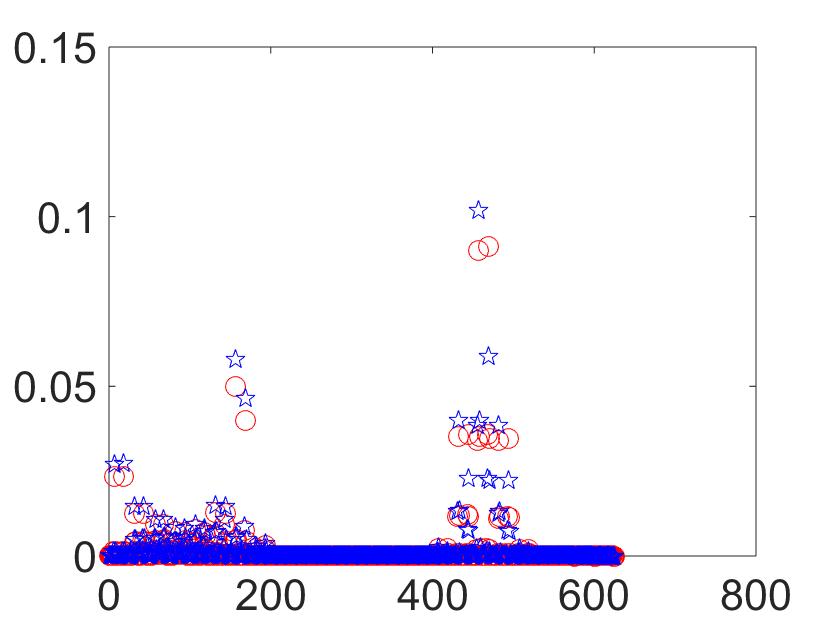}
\caption{Probabilities of occupation for the run in Fig.\ \ref{f.09}. The red circles are the probabilities for $\Rorig$, while the purple stars refer to $R$\@. The match is good, but not perfect. \label{f.10}\labeld{f.10}}
\end{figure}

Finally we allow a current in $\Rorig$ and look to what extent this is reproduced in $R$\@. For a path of $10^8$ steps things work out beautifully. See Fig.\ \ref{f.08}. However, this is longer than most of the paths I have used. For shorter paths things do not look so good. In Fig.\ \ref{f.09} I show the currents for a path of 1/10$^\mathrm{th}$ the length. Often the match is not nearly so good. On the other hand, a less sensitive quantity, the stationary state, comes out fairly well. In Fig.\ \ref{f.10} I show the (stationary) probability distributions, with that of $\Rorig$ shown as circles and for $R$ shown as stars.

\section{Discussion \label{s.3}\labeld{s.3}}

The technique described here reverses the usual application of ergodicity by deducing ``spatial'' behavior from a time series. The time series, the path taken by the Markov process, is used to reconstruct a matrix that would lead to that path. The method is simply to count the number of occurrences of each transition and to weight the estimator transition matrix accordingly. One fake step is added (from the final state to the initial state) to make the matrix irreducible.  The ergodicity is not always complete: the path may explore only a subset of the state space, hindered perhaps by some sort of barrier, effectively creating another ``constant of the motion'' (which precludes the usual ergodicity). Nevertheless, \textit{within} that subset information can be gleaned, as demonstrated above.

The method has been used in various applications to be reported elsewhere (including~\cite{mwthesis}).

In this discussion I want to consider situations where the method does \textit{not} work. The main drawback is the large values of $T$ needed to assure accuracy. One does not always have a sufficiently long path, compared to overall matrix size, $N$, which one knows. (Coarse graining can improve this, reducing $N$\@. However, further further familiarity with the system may be required.)

Consider also the effect of the choice of initial state, $x(1)$\@. For some paths this might be an unlikely state (for the true transition matrix or for some sub-matrix). This could mean that our method of making the matrix irreducible, namely adding a step from the final state to the initial state could seriously distort the transitions probabilities. In this case, it might pay to drop some of the initial wandering of the random walk and begin the tally only after the system has entered a region of higher probability. In such a situation the irreducibility addition could be less significant than the reduction in path length. Of course one cannot be sure that this accomplishes its intent, so that like much of physics \cite{ergodicityreasonable} one must play things by ear~\footnote{\noteear}.

One should also bear in mind the existence of metastable states. The path $x(t)$ may very well remain in a single metastable state throughout the period of observation. In this case, should one construct an embedding (\`a la the observable representation) that embedding \textit{only} represents the portion of the full transition matrix associated with this metastable state.

Currents are another issue. They lie at the heart of non-equilibrium phenomena and complexity \cite{ronchi}. Interestingly, they are also intimately tied to dissipation \cite{creation} and many of their properties have been studied in detail \cite{currentsnonequilibrium}. We illustrated a path that gave the probabilities more or less correctly, but for which the currents are inaccurate. That could be cured by using a yet longer path. Nevertheless, caution should be observed when using this method for more delicate quantities (such as current). In an article that uses techniques similar to those described here, Battle et al. \cite{battle} demonstrate the non-equilibrium nature of certain biological processes. Since generically a zero-current transition matrix develops currents when approximated, there is a need for long paths to be certain of establishing non-equilibrium properties.

Finally, it may happen that an experimenter has more than one path to work with. In that case my recommendation would be to attach one path to another (i.e., if you have $x_i(t)$, $1\le t\le T_i$ and $i=1,\dots,m$, then the total path $y(t)$ would have (e.g.) $y(T_1)=x_1(T_1)$ and $y(T_1+1)=x_2(1)$\@. This will have several false transitions but if the overall length is large they should not be important.

\section*{Acknowledgement}

The author is grateful to B. Gaveau, M. G. E. da Luz and M. E. Wosniack for helpful discussions. He also thanks the Brazilian granting agency, Ci\^encia Sem Fronteiras-CNPq (project 402193/2012-1), for financial support.

\appendix
\section{The observable representation \label{s.observable} \labeld{s.observable}}

There is a space $X$ of finite cardinality, $|X|=N<\infty$\@. A Markov process is defined on $X$, with transition probabilities
\be
R(x,y)=\Pr(x\leftarrow y)
\,,
\label{e.a010}
\ee
\labeld{e.a010}
for $x,y\in X$, in unit time step. Probability is conserved, implying (the matrix left eigenvector equation) $A_0R=A_0$, with $A_0(x)\equiv1$ and $A_0$ considered to be a row vector. We assume that $R$ is irreducible, implying the existence of a unique right eigenvector $Rp_0=p_0$, which is the stationary state of the Markov process. Under these conditions $p_0$ is strictly positive and normalized by $A_0 p_0=1$\@. The eigenvalues of $R$ are ordered by $1\equiv\lambda_0>|\lambda_1|\ge|\lambda_2|\ge\dots$, with an appropriate rule for complex $\lambda$\@. The \textit{left} eigenvectors of $R$ are written $A_0, A_1,\dots$, and are used to define the observable representation. The level-$m$ observable representation is the subset of Euclidean $m$-space defined by $(A_1(x),A_2(x)\dots,A_m(x))$ for all $x\in X$ and for real $A$\@. (When some $A_k$, for $k\le m$, is complex, other constructions are used.)

This simple construction can lead to insight on the dynamics taking place on $X$\@. See \cite{note:visualizing} and the references given there. For the present article we mention that this leads to a definition of phase transitions that allows for the existence of metastable states (which conventional analytic function theory does not). In particular when the first $m$ $\lambda$'s (after $\lambda_0$) are close to 1, with subsequent eigenvalues much smaller, there will be $m+1$ phases, with the figure in the observable representation taking the form of a simplex with $m+1$ vertices, each vertex a phase. This is what is reproduced in the body of this article. The phenomenon described is a generalization of a first order phase transition. It is a generalization because the usual theory (based on analyticity) lacks a description of metastability. And it is first order because of the gap (after $\lambda_m$ in eigenvalue values). Spectra without a gap were considered in \cite{imaging}.

Also mentioned in this article are \textit{currents}, defined as $J(x,y)=R(x,y)p_0(y)-R(y,x)p_0(x)$\@. Nonzero currents correspond to a breakdown in detailed balance and are characteristic of stationary states that are not equilibrium.


\end{document}